\documentclass[12pt]{article}
\usepackage[all]{xy}

\usepackage{etex}
\usepackage{amsmath,amssymb,amscd,bm,mathtools,array}

\usepackage[margin=1.75cm]{caption} 
\usepackage[labelfont=it,textfont={small,it}]{caption} 

\usepackage{color}
\usepackage[colorlinks=true, pdfstartview=FitV, linkcolor=blue, citecolor=blue, urlcolor=blue]{hyperref}

\usepackage[cyr]{aeguill}     
\usepackage[pdftex]{graphicx} 
\DeclareGraphicsExtensions{.jpg,.mps,.pdf,.png} 

\let\ssection=\section
\renewcommand{\section}{\setcounter{equation}{0}\ssection}

\newcommand{\bone}{{\bf 1}}

\newcommand{\balpha}{\boldsymbol{\alpha}}

\newcommand{\cE}{{\mathcal{E}}}

\newcommand{\rg}{\mathrm{g}}

\newcommand{\cL}{{\mathcal{L}}}

\newcommand{\bomega}{\boldsymbol{\omega}}

\newcommand{\barP}{\overline{P}}

\newcommand{\bp}{{\mathbf{p}}}
\newcommand{\hbp}{\bp}
\newcommand{\hbpe}{\bp_e}
\newcommand{\np}{\Vert\bp\Vert}
\newcommand{\nbp}{\Vert\bp\Vert}
\newcommand{\nhpe}{\Vert\hbpe\Vert}
\newcommand{\nhp}{\Vert\hbp\Vert}
\newcommand{\Pf}{\mathrm{Pf}}
\newcommand{\dP}{\dot{P}}
\newcommand{\cP}{{\mathcal{P}}}

\newcommand{\bbR}{\mathbb{R}}

\newcommand{\bs}{{\mathbf{s}}}
\newcommand{\ns}{\Vert\bs\Vert}
\newcommand{\cS}{{\mathcal{S}}}

\newcommand{\dS}{\dot{S}}

\newcommand{\Tr}{\mathrm{Tr}}

\newcommand{\hW}{W}

\newcommand{\dX}{\dot{X}}
\newcommand{\bx}{{\mathbf{x}}}

\newcommand{\half}{\frac{1}{2}}

\newcommand{\blue}[1]{{\textcolor{blue}{#1}}}

\newcommand{\pa}{\partial}
\newcommand{\lb}{\left[}
\newcommand{\rb}{\right]}
\newcommand{\lp}{\left(}
\newcommand{\rp}{\right)}
\newcommand{\bb}{\begin{eqnarray}}
\newcommand{\ee}{\end{eqnarray}}
\newcommand{\eee}{\nonumber\end{eqnarray}}
\newcommand{\qq}{\quad}
\newcommand{\dpp}{\vcentcolon}

\def\s{\sigma}

\begin{document}
\thispagestyle{empty}

\begin{center}
${}$
\vspace{3cm}

{\Large\textbf{Gravitational birefringence and an exotic formula for redshift}} \\

\vspace{2cm}

{\large
Christian Duval\footnote{
Aix-Marseille University, Universit\'e de Toulon, CPT, Marseille, France
\\\indent\qq supported by the ARCHIMED Labex (ANR-11-LABX-0060) funded by the
"Investissements d'Avenir" 
\\\indent\qq
French government program
\\\indent\qq
christian.duval@cpt.univ-mrs.fr },
Johanna Pasquet\footnote{Aix-Marseille University, CPPM, Marseille, France
\\\indent\qq supported by the OCEVU Labex (ANR-11-LABX-0060) funded by the
"Investissements d'Avenir" 
\\\indent\qq
French government program
\\\indent\qq
pasquet@cppm.in2p3.fr },
Thomas Sch\"ucker\footnote{
 Aix-Marseille University, Universit\'e de Toulon, CPT, Marseille, France
\\\indent\qq supported by the ARCHIMED Labex (ANR-11-LABX-0060) funded by the
"Investissements d'Avenir" 
\\\indent\qq
French government program
\\\indent\qq
thomas.schucker@gmail.com }
and
Andr\'e Tilquin\footnote{Aix-Marseille University, CPPM, Marseille, France
\\\indent\qq supported by the OCEVU Labex (ANR-11-LABX-0060) funded by the
"Investissements d'Avenir" 
\\\indent\qq
French government program
\\\indent\qq
 tilquin@cppm.in2p3.fr }

}

\vspace{2cm}

{\large\textbf{Abstract}}

\end{center}

We compute the birefringence of light in curved Robertson-Walker spacetimes and propose an exotic formula for redshift based on the internal structure of the spinning photon. We then use the Hubble diagram of supernovae to test this formula.

\vspace{2cm}

1802.09295

\eject

\section{Introduction}

Birefringence of light in absence of anisotropic material but in presence of a reflecting surface, was predicted by Federov \cite{Fed55} in 1955 and by Imbert \cite{Imb72} in 1972. It implies an offset of the order of a wavelength between the outgoing photons of either circular polarisation. In 2008 this was indeed observed \cite{BNKH08,HK08}. 

Birefringence of light in presence of the gravitational field of the Schwarzschild metric was computed by Saturnini \cite{Sat76} in 1976 and recently by two of us \cite{k0} in a flat Robertson-Walker metric. 

In the second case, the photons follow helical trajectories whose chirality is given by the two circular polarisations. The distance between the two trajectories at arrival is -- as in the Fedorov-Imbert effect -- of the order of a wavelength but direct observation of this offset seems much more difficult.  

The period of the helix is equal to the period of spin precession and depends on the atomic period of the photon. But it also depends on the acceleration of the universe. This 'internal structure' of the spinning photon encourages us to propose an exotic definition of redshift based on the precession period rather than on the atomic period.

The aim of this paper is two-fold. First, we generalise our computations of reference \cite{k0}  from flat to curved Robertson-Walker metrics, sections \ref{theory} and \ref{perturb}, and present the resulting exotic definition of redshift, section \ref{exot}.
Second, in a universe filled with cold matter only, we confront our exotic definition to the Hubble diagram of supernovae, section \ref{observe} with the initial hope that the acceleration dependence in the exotic redshift might yield a good fit with less (dark) matter. 

\section{Theory}\label{theory}

Already for spinless, point-like particles in general relativity, the massless limit is delicate. It cannot be continuous because, by the equivalence principle, the geodesics of particles with positive mass do not depend on this positive mass. The point-like limit itself is delicate because of the non-linear nature of general relativity. Nevertheless WKB methods do apply under certain conditions and somehow reconcile waves and localized particles. 

Spin of massless particles invokes quantum physics with its known problems in general relativity and we should not be surprised to face additional delicate limits. Indeed the limit of flat space-time and the limit of vanishing spin are both ill defined in general. Nevertheless WKB approximations still produce, in particular cases, the same trajectories as obtained through the geometric approach. Despite these examples, we are still lacking a derivation of the Souriau-Saturnini equations from Maxwell's theory coupled to general relativity in an appropriate approximation. Without such a derivation, these equations are merely a coordinate independent model describing the propagation of photons through a gravitational field and hopefully accessible to observation.

The first chapter of our reference \cite{k0} introduces the reader to the vast literature of the massless particle with spin in its different approaches and the appendix of reference \cite{k0} presents a self-contained derivation of the Souriau-Saturnini equations from the Mathisson-Papapetrou-Dixon equations.

Consider the worldline $X(\tau)$ of a massless particle with non-vanishing spin in a gravitational field described by a metric $\rg$. Denote by $P(\tau)$ the 4-momentum of the particle and by $S(\tau)$ its spin tensor (viewed as a linear map). The worldline is determined by the Souriau-Saturnini equations \cite{Kun72,DFS72,Sou74,Sat76}, which are obtained from the Mathisson-Papapetrou-Dixon equations \cite{Mat37,Papa51,Tau64,Dix70,DFS72,Sou74} by setting $SP=0$. This constraint implies that $P^2$ and the `scalar spin' $s$ defined by $s^2\dpp=-{\textstyle\frac{1}{2}} {\rm Tr}(S^2)$ are both constants of motion. For the photon of course we set:
\bb
P^2=0\,,\qq\qq\qq s=\pm\hbar\,.
\ee
  Then the Souriau-Saturnini equations read:
\begin{eqnarray}
\label{dotXter}
\dX&=&P+\frac{2}{R(S)(S)}S R(S) P\,,\\[4pt]
\label{dotPter}
\dP&=&-s\,\frac{\Pf(R(S))}{R(S)(S)}\,P\,,\\[4pt]
\label{dotSter}
\dS&=&P\overline{\dX}-\dX\barP.
\end{eqnarray}
Our conventions are: $\dot X^\mu\dpp=dx^\mu/ d\tau$, ($\mu=1\,,2\,,3\,,4$) and $(\overline P)_\nu\dpp= g_{\nu \mu} P^\mu$. Over $P$ and $S$, the dot denotes the covariant derivative with respect to $\tau$. 
 $R(S)(S)\dpp=-\Tr(SR(S))\dpp=R_{\mu\nu\alpha\beta}\,S^{\mu\nu}S^{\alpha\beta}$. Of course we must assume $R(S)(S)\neq0$. In (\ref{dotPter}), the Pfaffian of the skewsymmetric linear map  $F=R(S)$ is such that $\star(F)F=\Pf(F)\,\bone$, where~$\star$ is the Hodge star; we have $\det(F)=-\Pf(F)^2$. Note that $\Pf(F)=-\frac{1}{8}\sqrt{-\det(\rg_{\alpha\beta})}\,\varepsilon_{\mu\nu\rho\sigma}F^{\mu\nu}F^{\rho\sigma}$ with $\varepsilon_{\mu\nu\rho\sigma}$ the Levi-Civita symbol such that $\varepsilon_{1234}=1$. 
 
The Souriau-Saturnini equations also imply that 4-momentum and 4-velocity are orthogonal, $\overline P\dot X=0$.  

Let us write the Robertson-Walker metric with respect to the Euclidean coordinates $\bx$  of $\bbR^3$ after stereographic projection and with respect to cosmic time $t$:
\bb
\rg = -a(t)^2\,\frac{\Vert{}d\bx\Vert^2}{b(\bx)^2}+dt^2
\quad
{\rm with}
\quad
b(\bx)\dpp=1+\frac{K}{4}\Vert\bx\Vert^2. \label{rw}
\label{gbis}
\ee
Our conventions are such that the curvature scalar of 3-space is $6K$.
The Euclidean coordinates will come in handy because of frequent vector products.

%
%

In these coordinates we write the momentum of the particle as
\begin{equation}
P=\left(
\begin{array}{c}
\displaystyle
\frac{b}{a}\hbp\\[8pt]
\displaystyle
\nhp
\end{array}\right)
\label{Pbis}
\end{equation}
with 3-momentum $\hbp$ and {\it positive} energy $\nhp\dpp=\sqrt{\hbp\cdot\hbp}$. Accordingly, we write the spin $(1,1)$-tensor as
\begin{equation}
S=\left(
\begin{array}{cc}
j(\bs)&\displaystyle
-\frac{(\bs\times\hbp)}{\nhp}\frac{b}{a}\\[6pt]
\displaystyle
-\frac{(\bs\times\hbp)^T}{\nhp}\frac{a}{b}&0
\end{array}\right)\,.
\label{Sbis}
\end{equation}
 Transposition is denoted by a superscript $\cdot^T$ and we use $j(\bs)\!:\hbp\mapsto\bs\times\hbp$. The relation between the scalar spin and the spin vector $\bs\in\bbR^3\setminus\{0\}$ then reads
\begin{equation}
s=\frac{\bs\cdot\hbp}{\nhp}
\label{sbis}
\end{equation}
and we find
\begin{equation}
R(S)=-\frac{2}{a^2}\left(
\begin{array}{cc}
\displaystyle
(K+a'^2)j(\bs)&
\displaystyle
-\frac{\bs\times\hbp}{\nhp}\,b\,a''\\[6pt]
\displaystyle
-\frac{(\bs\times\hbp)^T}{\nhp}\,\frac{a^2a''}{b}&0
\end{array}\right)\,,
\label{RSbis}
\end{equation}
together with $\det(R(S))=0$ implying that the 4-momentum $P$ is  parallel-transported.

Furthermore we have
\begin{equation}
R(S)(S)=\frac{4}{a^2}\left(\Vert\bs\Vert^2(aa''-(K+a'^2))-s^2\,aa''\right).
\label{RSSbis}
\end{equation}
and
\begin{equation}
S R(S)P=\frac{2}{a^2}((K+a'^2)-aa'')\left(\Vert\bs\Vert^2 P-s\,\hW\right)
\label{SRSPbis}
\end{equation}
with the `Pauli-Lubanski' vector
\begin{equation}
\hW\dpp=\nhp\left(
\begin{array}{c}
\displaystyle
\frac{b}{a}\bs
\\[8pt]
\displaystyle
s
\end{array}\right)\,
\label{Wbis}
\end{equation}
interpreted as the \textit{polarization vector} of the massless particle in the gravitational field. 

Next we trade the curve parameter $\tau$ for the cosmic time. The fourth component of Equation (\ref{dotXter}),  
\begin{equation}
\frac{dt}{\!d\tau}=-\frac{4s^2\nhp}{R(S)(S)}\frac{(K+a'^2)}{a^2}\,,
\label{dt/dtau}
\end{equation}
tells us that we must assume $R(S)(S)\neq0$, as already noted, but also $K+a'^2\neq0$.
Then we can write
\begin{equation}
\frac{dX}{\!dt}=\frac{aa''}{\nhp(K+a'^2)}\left[P-\left(1-\frac{K+a'^2}{aa''}\right)\frac{\hW}{s}\right]\,,
\label{dX/dt}
\end{equation}
where the Pauli-Lubanski vector $\hW$ features the polarization-driven `anomalous velocity'. 

Let us write the {\it de}celeration parameter modified by curvature as \bb
Q(t)\dpp=-a(t)a''(t)/(K+a'(t)^2)\,.
\ee
 Then the equations of motion read in our $3+1$ decomposition:
\goodbreak
\begin{eqnarray}
\label{dxBis}
\frac{d\bx}{dt}
&=&
\frac{b}{a}\lb{}-Q\frac{\hbp}{\nhp}+\left(1+Q\right)\frac{\bs}{s}\rb,\\[8pt]
\label{dpBis}
\nonumber
\frac{d\hbp}{dt}
&=&
-\frac{a'}{a}\left[-Q\,\hbp+\nhp\left(1+Q\right)\frac{\bs}{s}\right]\\[8pt]
&&
+\frac{K}{2a}\left[\left(1+Q\right)
(\bp\cdot\bx)\frac{\bs}{s}
-Q\frac{(\bp\cdot\bx)}{\nhp}\,\bp-\np\,\bx \rb,\\[8pt]
\label{dsBis}
\nonumber
\frac{d\bs}{dt}
&=&
-\left(1+Q\right)\frac{\bs}{s}\times\hbp
-\frac{a'}{a}\lb\,\bs
+\lb{}s\,Q-\left(1+Q\right)\frac{\ns^2}{s}\rb\frac{\hbp}{\nhp}\rb\\[8pt]
&&
+\frac{K}{2a}\lb(1+Q)\frac{\bs}{s}\times\left(\bs\times\bx\right)
-\,Q \lb(\bs\!\cdot\!\bx)\frac{\hbp}{\nhp}-s\,\bx\rb\rb. \label{ds/dt}
\end{eqnarray}
With
$
b_\pm\dpp=1\pm\frac{K}{4}\Vert\bx\Vert^2
$ 
(implying $b=b_+$) and defining the transverse spin
\begin{equation}
\bs^\perp\dpp=\bs-s\frac{\bp}{\nbp}\,,
\label{sperp}
\end{equation} 
we have the following constants of motion:
\begin{eqnarray}
\nonumber
\cP
&{=}&
{\frac{b_-}{b_+}\left[a\,\bp+a'\bs\times\frac{\bp}{\nbp}\right]
}\\[8pt]
&&
{+\frac{K}{2b_+}\left[2\,\bx\times\bs+\lb{}a'\,\bx\cdot\left(\bs\times\frac{\bp}{\nbp}\right)+a\,(\bx\cdot\bp)\rb\bx\right]\,,
}\label{calP}\\[8pt]
{\cL}
&{=}&
{\frac{1}{b_-}\left[\bx\times\cP+b_+\,\bs-\frac{K}{2}(\bs\cdot\bx)\,\bx\right]\,,
}\\[8pt]
\cE&=&a\,\nbp\,,
\label{cE}
\\[8pt]
s&=&\frac{\bs\cdot\bp}{\nbp}\,,
\label{scalarspin}\\[8pt]
\cS&=&\sqrt{K+a'^2}\,\Vert\bs^\perp\Vert\,.\label{cS}
\end{eqnarray}
As detailed for the flat case in reference \cite{k0} the conservation of momentum $\cP$ and angular momentum $\cL$  follows from the invariance of the Robertson-Walker metric (\ref{rw}) under infinitesimal ``translations'' and rotations
\bb
\delta\bx=\bomega\times\bx+\balpha\,b_-+\frac{K}{2}\bx(\bx\cdot\balpha)
\ee
with $\balpha,\bomega\in\bbR^3$. The conservation of energy $\cE$ comes from the {\it conformal} Killing vector $a(t)\,\pa/\pa t$.
Generalizing the proof of the conservation of transverse spin $\cS$ from \cite{k0} to the curved case is difficult and we give an alternative one, starting with
\bb
\,\frac{d}{dt}\, \lp\half || \bs ||^2\rp 
= 
\,\frac{d}{dt}\, \lp\half || \bs^\perp ||^2\rp 
=
\bs \cdot \,\frac{d\bs}{dt}\, 
=
Q \,\frac{a'}{a}\,  || \bs^\perp ||^2.
\ee
Here we have used the equation of motion for the spin vector (\ref{ds/dt}) and the relation between the scalar spin and the spin vector (\ref{sbis}). With the definition of the deceleration parameter, 
 $Q = -a a''/(K+a'^2)$, we finally obtain
$ d/dt[(K+a'^2) || \bs^\perp ||^2] = 0 $.

As already in the flat case \cite{k0} we can use the constants of motion to express the spin vector as a function of momentum and position:
\bb
\bs=\frac{1}{b_+}\left[b_-\,\cL+\frac{K}{2}(\bx\cdot\cL)\,\bx-\bx\times\cP\right].\label{spin}
\ee
However the analogous simple expression for the momentum, that we enjoyed in the flat case, equation (4.4) of reference \cite{k0}, eludes us for non-vanishing curvature $K$.

Using the constants of motion we can rewrite Equation (\ref{RSSbis}):
\begin{equation}
R(S)(S)=-\frac{4}{a^2}\left(\Vert\cS\Vert^2(1+Q)+s^2\,(K+a'^2)\right)\,,\label{RSS}
\end{equation}
which must never vanish. 

Let us anticipate that later on we will have to assume that $1+Q$ alone never vanishes. Therefore we need to verify both constraints $1+Q>0$ and $K+a'^2>0$ independently at any time. They can be recast conveniently using the Friedman equations in presence of a single matter component with energy density $\rho $ and with vanishing pressure. With the usual dimensionless cosmological parameters,
\bb
\Omega _\Lambda\, \dpp=\,\frac{\Lambda }{3H^2}\,,
\qq  
\Omega _k\, \dpp=\,\frac{-K }{a^2H^2}\,,
\qq
\Omega _m\, \dpp=\,\frac{8\pi G\,\rho  }{3H^2}\,,
\ee
we find that $1+Q>0$ if and only if $\Omega _\Lambda +\Omega _k<1$ or equivalently $\Omega _m>0$ and $K+a'^2>0$ if and only if $\Omega _k<1$ or equivalently  $\Omega _m+\Omega _\Lambda >0$.

\section{Perturbation}\label{perturb}
We would like to solve the equations of motion (\ref{dxBis} - \ref{dsBis}) with initial conditions
 at $t=t_e$, the time of emission:
\bb \bx_e=0,\qq
\hbp_e=\begin{pmatrix}
\nhpe\\0\\0
\end{pmatrix},\qq
\bs_e=\begin{pmatrix}
s\\s^\perp_e\\0
\end{pmatrix}
\label{initial}
\ee
with $s_e^\perp\dpp=\Vert\bs_e^\perp\Vert\geq0$.
For `enslaved spin', $s_e^\perp=0$, we retrieve the null geodesics: 
$x^1=\tilde x$, $p_1=a(t_e)/a(t)\,\nhpe$, $s_1=s$ with
\bb
 \tilde x(t)\dpp =
 \left\{ \begin{array}{ll}
 2/\sqrt{|K|}\, \tan\lp\sqrt{|K|}/2\,\int_{t_e}^td\tilde t/{a(\tilde t)}\rp   & K >0\\[3mm]
\int_{t_e}^td\tilde t/{a(\tilde t)}&K =0\\[3mm]
 2/\sqrt{|K|}\, \tanh\lp\sqrt{|K|}/2\,\int_{t_e}^td\tilde t/{a(\tilde t)}\rp     & K <0
 \end{array}\right. 
 \ee
 and the six other components vanish. 
 
 We will be dealing with two small parameters, typically of the order of $10^{-33}$,
\bb \eta\dpp =\,\frac{s}{\cE}\,  ,\qq\qq
\epsilon\dpp =\,\frac{s^\perp_e}{\cE}
\label{etaepsilon}
\ee
We consider $\eta$ to be a fixed, non-zero number and $\epsilon$ to vary between 0 and $|\eta|$. Indeed we know that for $\epsilon =0$ our trajectory is the null geodesic and we want to know how the trajectory of the photon deviates from this geodesic to first order in $\epsilon$. At the end of our calculation we will put $\epsilon =| \eta|$ for the photon.

From numerical solutions in the flat case $K=0$, we know that the six  components, which vanish for vanishing $\epsilon$, are at least of first order in $\epsilon$ justifying the Ansatz:
\bb \bx\,=\begin{pmatrix}
\tilde x\,+\!\!\!\!\!&\epsilon y_1\\&\epsilon y_2\\&\epsilon y_3
\end{pmatrix},\qq
\,\frac{\hbp}{\cE}\, =\begin{pmatrix}
1/a\,+\!\!\!\!\!&\epsilon q_1\\&\epsilon q_2\\&\epsilon q_3
\end{pmatrix},\qq
\,\frac{\bs}{\cE}\, =\begin{pmatrix}
\eta\,+\!\!\!\!\!&\epsilon r_1\\&\epsilon r_2\\&\epsilon r_3
\end{pmatrix}\,.
\label{ansatz}
\ee
Computing $\Vert\bp/\cE\Vert^2 $ we immediately find that $q_1\sim0$. Likewise we compute 
\bb s/\cE=\lp\frac{\bp}{\cE}\, \cdot\,\frac{\bs}{\cE}\rp/\,\frac{\Vert\bp\Vert}{\cE}\ee
with our Ansatz and find that $r_1\sim0$. Now consider the spin divided by $\cE$ written in terms of the constants of motion, equation (\ref{spin}), and replace the constants of motion by the initial values and replace $\bx$ by our Ansatz. Then we get to first order in $\epsilon$:
\begin{align}
&&&&r_2\sim&&\!\!\!\!\!\!\!\!\!\!\!\!\!\!\!\!\!\!\frac{b_--a'_e\tilde x}{b_+}\, {-}&\,\frac{y_3}{b_+}\,,&&&&&
 \\[1mm]
&&&&r_3\sim &&&\,\frac{y_2}{b_+}\,.&&&&&
\end{align}
Finally we use equation (\ref{calP}) and write the vector of constants of motion $\cP/\cE$ for $t=t_e$ and for arbitrary $t$ using the Ansatz.  The first component is fulfilled identically to first order. The other two components yield
\begin{align}
q_2\sim\,&&&\lb-\frac{a'}{ab_+}+\frac{K\tilde x}{2ab_-}\lp\frac{2}{b_+} -1\rp\rb y_2,\\[2mm]
q_3 \sim\,&{-}
\frac{a_e'b_+}{ab_-} {+}\frac{a'b_-}{ab_+} {-}\frac{a_e'a'\tilde x}{ab_+} {-}\frac{K\tilde x}{ab_+} 
{+}\frac{a_e'K\tilde x^2}{ab_+b_-}\!\!\!\!\!\!\!\!\!&
\!\!\!\!\!\!\!\!\!\!\!\!\!\!\!\!\!\!\!\!\!\!\!\!\!\!+&\lb-\frac{a'}{ab_+}+\frac{K\tilde x}{2ab_-}\lp\frac{2}{b_+} -1\rp\rb y_3. 
\end{align}
We are now ready to linearize the 3-velocity (\ref{dxBis}):
\begin{align}
\,\frac{dx_1}{dt}\,\sim &\,\frac{b}{a}\, \qq\qq\ {\rm implying}\qq\qq\  {\,\frac{dy_1}{dt}\,=0 ,}\\[2mm]
\,\frac{dy_2}{dt}\, \sim &-Q\,b\,q_2+\,\frac{1+Q}{a}\, b\,\frac{1}{\eta}\, r_2\,\sim\,\frac{1+Q}{\eta a}\, [{-}y_3+b_--a'_e\tilde x]\,,\\[2mm]
\,\frac{dy_3}{dt}\, \sim &-Q\,b\,q_3+\,\frac{1+Q}{a}\, b\,\frac{1}{\eta}\, r_3\,\sim\,\frac{1+Q}{\eta a}\, [{+}y_2]\,.
\end{align}
{Since $y_1(t_e)=0$, we conclude that $y_1$ vanishes everywhere.} Note that the momenta $q_2$ and $q_3$ do not contribute in leading order. Let us suppose that $1+Q$ does not vanish between emission time $t_e$ and today $t_0$. Then we may define a new time coordinate $\theta $ by:
\bb
\,\frac{d\theta }{dt}\, =\,\frac{1}{{|\eta|}}\,\frac{1+Q}{a}\, \qq\qq {\rm and}\qq\qq \theta (t_e)=0,
\ee
and new dependent variables $\tilde z(\theta)\dpp=\tilde x(t(\theta ))$, $z_2(\theta )\dpp =y_2(t(\theta ))$ and $z_3(\theta )\dpp =y_3(t(\theta ))$. Then we have:
\begin{align}
\,\frac{dz_2}{d\theta }\, &\sim\,{{\rm sign}(\eta)}[{-}z_3+b_--a'_e\,\tilde z],\\[2mm]
\,\frac{dz_3}{d\theta }\, &\sim\, {{\rm sign}(\eta)}\,z_2\,.
\end{align}
Setting $\epsilon=|\eta|=T_e/(2\pi \,a_e)$, the solution is $z_2\,\sim\,{\rm sign}(\eta)\,\sin\theta $ and  $z_3\,\sim\,{-}\cos\theta -1+a_e'\tilde z$, a helix of constant period $2\pi$ with respect to the time coordinate $\theta $.   The period is variable with respect to cosmic time:
\bb 
T_{\rm helix}(t)\,\sim \,\frac{a(t)}{a_e}\,\frac{1}{1+Q(t)}\, T_e,
\ee
where  $T_e$ is the atomic period of the light (spin 1) at emission. The radius of the helix is also time dependent:
\bb 
R_{\rm helix}(t)\,\sim \,\frac{a(t)}{a_e}\,\frac{1}{1+K\tilde x(t)^2}\, \lambda_e ,
\ee
where  $\lambda_e =c\,T_e$ is the wavelength of the light at emission.
To leading order, the center of the helix has comoving coordinates 
\bb
\begin{pmatrix}
 \tilde x(t)\\ 0 \\ T_e/(2\pi \,a_e)\,\lb1-a'_e\,\tilde x(t)-K\tilde x(t)^2\rb
 \end{pmatrix}
.\ee

\section{An exotic definition of redshift}\label{exot}

Taking due account of its spin, the photon propagates through a Robertson-Walker universe on a helix.

Of course the main question is whether the offset between two helices of opposite polarisation, 'birefringence', is observable today.  The offset is to oscillate between 0 and $2\,R_{\rm helix}(t_0)$ with a period of  ${\textstyle\frac{1}{2}} T_{\rm helix}(t_0)$, where $ T_{\rm helix}(t)$ is the period of the helix and at the same time it is the period of precession of the spin vector $\bs$ around its direction of mean propagation.

\smallskip
Thanks to its spin, the photon carries two informations (besides its direction): 
\begin{itemize}\item
The first information is the well measured {\bf atomic period} today $T_0$, which is related to the atomic period at emission $T_e$ by
\bb T_0\,=\,\frac{a(t_0)}{a(t_e)}\, T_e.\ee
This relation has two derivations: In the classical one, we compute   the cosmic times of flight of a photon emitted by a co-moving source at time $t_e$ and arriving at a co-moving observer at time $t_0$ and of a second photon emitted at time $t_e+T_e$ by the same co-moving source and arriving at the same co-moving observer at time $t_0+T_0$. We note that cosmic time is equal to proper time for both source and observer. The second derivation is quantum, as it uses de Broglie's relation between the atomic period $T_e$ and the energy $\Vert\bp_e\Vert$ of the emitted photon. It also uses the conserved `energy' $\cE=a\Vert\bp_e\Vert$, Equation (\ref{cE}). 

This harmony between de Broglie's relations and general relativity has  been verified experimentally for the first time in 1960 in the (static) gravitational field of the Earth at Harvard by Pound and Rebka \cite{pr}.
\item
The photon carries a second information, its {\bf period of precession} today $T_{\rm helix}(t_0)=\dpp\,T_{{\rm helix}\,0}$ which is related to its atomic period $T_0$ today by 
\bb T_{{\rm helix}\,0}\,=\,\frac{1}{1+Q(t_0)}\, T_0\,.\label{bold}\ee
The derivation, presented above, also involves the conserved `angular momentum' $\cL$ and `spin' $\cS$ and $s$.
\end{itemize}

While waiting for a direct observation of birefringence, Equation (\ref{bold}) invites us to  be bold and (ignoring the mentioned harmony) assume that,  when telling us their redshift,  photons use the second information they carry, $z=(T_{{\rm helix}\,0}-T_{{\rm helix}\,e})/T_{{\rm helix}\,e}  $, and not the first one, $z=(T_{0}-T_{e})/T_{e}  $, which we must assume when we ignore spin.  

This assumption leads to an exotic formula for the redshift,
\bb
 z+1\,=\,\frac{a(t_0)}{a(t_e)}\,\frac{1+Q(t_e)}{1+Q(t_0)}\,,\label{exotic}
 \ee
which differs substantially from the standard formula by the presence of the modified deceleration para\-meter $Q$.  

Let us otherwise remain conservative and assume that gravity is well described by general relativity up to cosmic scales. We also assume the cosmological principle, i.e. that at cosmic scales our universe is maximally symmetric, and that supernovae of type Ia are standardizable candles. Then using again the conserved `energy' $\cE=a\Vert\bp_e\Vert$, the apparent luminosity is given by
\bb 
\ell \, = \,\frac{L}{4\pi\,a_0^2}\, \frac{1}{{\rm si}^2(t_0)}\, \frac{a_e^2}{a_0^2}\,, \label{lum}
\ee
where $L$ is the absolute luminosity of the supernova and 
\bb
 {\rm si}(t)\dpp =
 \left\{ \begin{array}{ll}
 1/\sqrt{|K|}\, \sin \lp\sqrt{|K|}\,\int_{t_e}^td\tilde t/{a(\tilde t)}\rp   & K >0\\[3mm]
\int_{t_e}^td\tilde t/{a(\tilde t)}&K =0\\[3mm]
 1/\sqrt{|K|}\, \sinh\lp\sqrt{|K|}\,\int_{t_e}^td\tilde t/{a(\tilde t)}\rp     & K <0
 \end{array}\right. \,.
 \ee
We are now ready to confront our exotic formula with the Hubble diagram of supernovae.

\section{Observation}\label{observe}

We use the 740 type Ia supernovae from the Joint Light Curve Analysis (JLA) \cite{jla}. 
The JLA published data provide the observed uncorrected peak magnitude 
($m_{\rm peak}$), the time stretching of the light-curve ($X1$) and color ($C$) at maximum
brightness due to intrinsic supernovae property and extinction by dust in the host galaxy.
These quantities are estimated in the restframe of each supernova by the SALT2 empirical model of 
Type Ia supernovae spectro-photometric evolution with time trained on the whole supernovae sample \cite{salt2}.

The reconstructed magnitude after the SALT2 fitting procedure reads:
\bb m_r = m_{\rm peak} + \alpha_s X1 - \beta_c C, \ee
where $\alpha_s$ and $\beta_c$ are global parameters fitted to the Hubble diagram simultaneously with all the other parameters.
The expected magnitude can be simply written as:
\bb m_e = m_s -2.5\,log_{10}\,\ell(a(t_e))\,, \ee
where  $\ell(a(t_e))$ is given by (\ref{lum}) and $m_s $ a global normalisation parameter. 

Because of the new relation between observed redshift and scale factor at emission (\ref{exotic})
the restframe of each supernovae is at a different emission time compared to the one given by the standard redshift. Thus, the global light
curve fit must be redone.

Using both Friedman equations for a non flat Universe, and setting $a(t_0)=1$ the new observed redshift reads:

\bb
 z+1\,=\,\frac{1}{a(t'_e)}\,\frac{\Omega _{m 0} + \Omega _{\Lambda 0}}{\Omega _{m 0} + \Omega _{\Lambda 0} \,a(t'_e)^{3}} \,,\label{zsolve}
\ee
where $t'_e$ is the new photon emission time and ($\Omega _{m 0}$, $\Omega _{\Lambda 0}$) the cosmological parameters today.

Consequently the scale factor at emission $a(t'_e)$ for a given observed redshift is now a function of the cosmological model.
 For a positive cosmological constant, the correction factor is always greater than one which implies that supernovae
appear to be closer to us.  
Notice that for a vanishing cosmological constant the exotic redshift is similar to the standard one.

We use a numerical gradient method to invert (\ref{zsolve}) with a minimum step in scale factor well below the associated experimental
redshift error. This new scale factor  $a(t'_e)$ is then used as input to the SALT2 public software through a new effective redshift
for each supernova:
 \bb z_{\rm eff}+1\dpp = 1/a(t'_e)\,. \ee

The SALT2 fitting procedure returns new values for $m_{\rm peak}$, $X1$ and $C$ as well as the full covariance matrix between those
parameters. Finally the JLA public likelihood software is used to compute the general  $\chi^2$ expressed in terms
of the full covariance matrix including correlations and systematics. It reads
\bb \chi^2 = \Delta M^T V^{-1} \Delta M, \ee
where  $\Delta M$  is the vector of differences between the expected supernova 
magnitudes $m_e$ and the reconstructed experimental magnitudes at maximum of
the light curve $m_r$. The minimisation of the $\chi^2$ is preformed  with the MINUIT minimizer software \cite{minuit}.

\begin{figure}[h]
\begin{center}
\includegraphics
[width=15cm, height=15cm]
{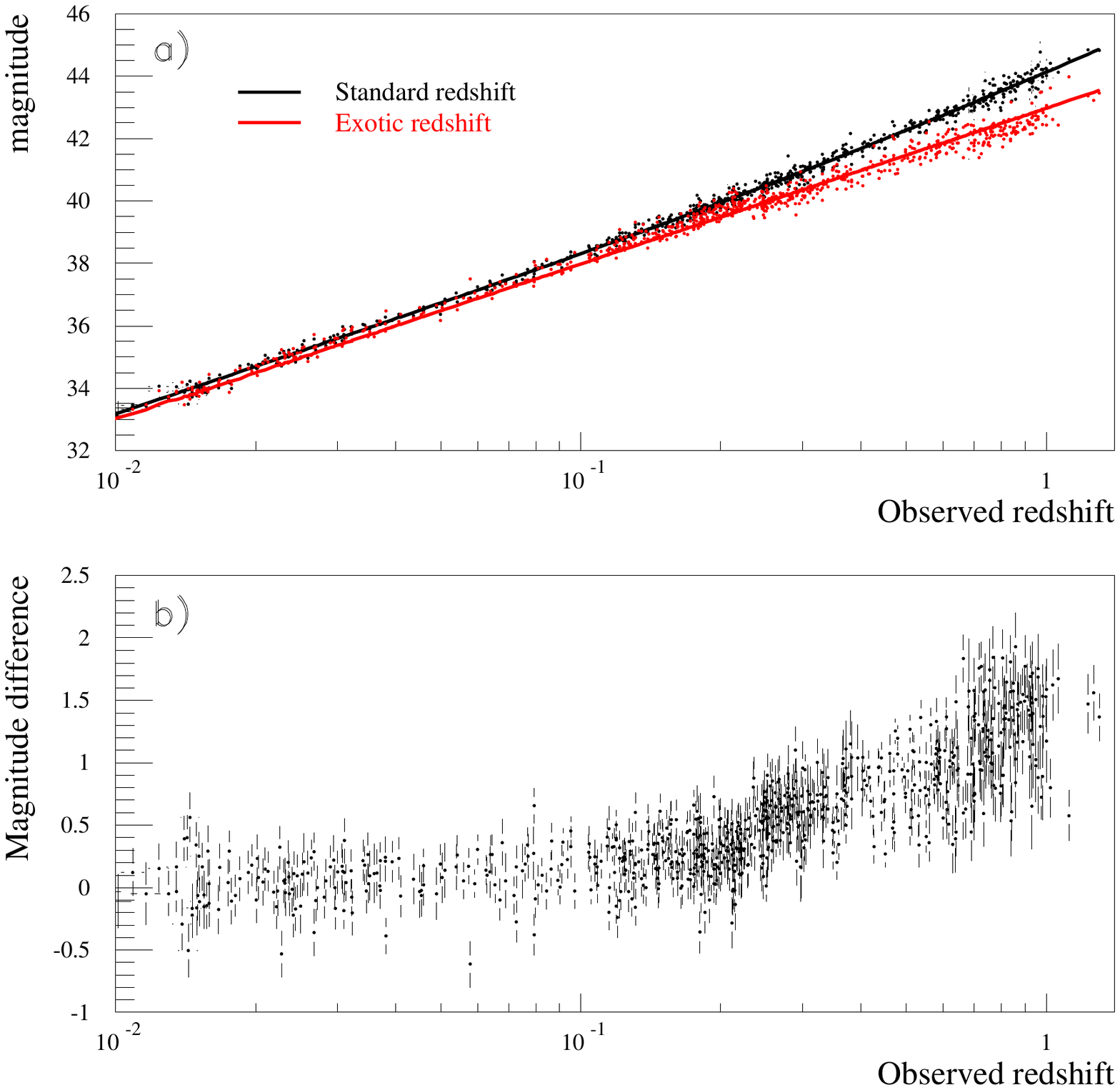}

\caption{a) Hubble diagram with standard redshift  in logarithmic scale (black points) and 
with exotic redshift (red points) at the same fiducial cosmology $\Omega_{m0} = 0.22$ and $\Omega_{\Lambda 0} = 0.63$
artificially calibrated at the same low redshift magnitude.
b) Relative difference in magnitude between standard and exotic redshifts after light curve re-fitting.}
\label{hubble}
\end{center}
\end{figure}

Figure \ref{hubble}-a shows the Hubble diagram for both redshift definitions. The black points are the reconstructed
magnitude at peak for the standard redshift
definition and the black curve corresponds to the global fit on all parameters ($m_s,\,\alpha_s,\,\beta_c,\,\Omega_{m0 },\,\Omega _{\Lambda 0}$). The red points are the reconstructed magnitudes given by the
SALT2 light curve refitting using the new redshift definition with the same standard fiducial cosmology and 
calibrated at the same low redshift value where both redshift definitions are identical. The red curve is obtained
by refitting only ($m_s,\,\alpha_s,\,\beta_c$). These two curves and the set of points show the important effect of the light curve
refitting on the reconstructed magnitude. This is better seen in figure \ref{hubble}-b which shows the reconstructed magnitude 
difference $m_r$(standard)$-m_r$(exotic) due to the light curve refitting with the exotic redshift.
We estimate the intrinsic magnitude dispersion as the square root of the mean square between curves and points: $\sigma_{intrinsic}$(standard)$=0.17$ and  $\sigma_{intrinsic}$(exotic)$=0.27$.
The standardization of supernova magnitudes with the exotic redshift is worse than with the standard one.

For completness, table \ref{table1} shows the results of the full fit with the standard redshift (line 1) and of the partial fit with the exotic redshift
assuming the same fiducial cosmology (line 2) corresponding respectively to black and red curves in figure \ref{hubble}-a.
The high $\chi^2$ value of line 2 is a consequence of the partial refit done at the same fiducial cosmology.
This is why we have to do a global refit (shown in line 3).

\begin{table}[htbp]
\begin{center}
\begin{tabular}{||c||c|c|c|c|c||} \hline
Redshift  &  $\alpha_s$      &   $\beta_c$    &    $\Omega_{m 0}$     &  $\Omega _{\Lambda 0}$ & $\chi^2$  \\  \hline
Standard  &$0.138 \pm 0.006 $&$3.14 \pm 0.08$&$0.22 \pm 0.11 $ &$0.63 \pm 0.16 $ & $748.9$  \\  \hline
Exotic (partial fit)    &$0.099 \pm 0.005 $&$1.58 \pm 0.02$&fixed to $0.22 $ & fixed to $0.63$ & $1609$  \\  \hline
Exotic (refit)    &$0.137 \pm 0.006 $&$3.16 \pm 0.08$&$-0.15 \pm 0.07 $ & $(-3 \pm 2) 10^{-4}$ & $757.6$  \\  \hline
\end{tabular}
\caption[]{1 $\sigma$ errors for stretch, color and cosmological parameter fits. Line 1 corresponds to the standard redshift,
line 2 to the exotic redshift after refitting light curves but at the same fiducial cosmology than line 1. Line 3 corresponds to the global fit using
exotic redshift and light curve refitting.}
\label{table1}
\end{center}
\end{table}

To find the true minimum with the exotic redshift we face a new problem: we cannot use the standard minimization method MINUIT due to numerical instabilities  in the light curve fit for all cosmologies. Indeed,  MINUIT requires 
first numerical derivatives of the $\chi^2$ with respect to $\Omega_{m 0}$ and $\Omega _{\Lambda 0}$ which lead to many fake minima.

To overcome this problem we prefer to explore the $\Omega_{m 0}$, $\Omega _{\Lambda 0}$ parameter
space using an accurate grid of points. At each point the SALT2 fit is performed on light curves using the effective
redshift and the output is used to minimize the  $\chi^2$ with respect to all other parameters using the JLA standard likelihood
and MINUIT. To avoid any a priori about the final result and despite the fact that both constraints $1+Q>0$ and $K+a'^2>0$ are
not fulfilled,
we decided to explore  $\Omega_{m 0}$ from $-1$ to 1 and $\Omega _{\Lambda 0}$ from $-1$ to 3
on a grid of about 30.000 points. This required three and a half CPU years made available to us by the Dark Energy Center 
\footnote{DEC or Dark Energy Center is a HPC cluster of 800 cores funded by the OCEVU Labex (ANR-11-LABX-0060)}. After identifying
the true minimum, we explored more accurately the $\chi^2$ in a smaller region in  $\Omega_{m 0}$, $\Omega _{\Lambda 0}$ to construct the
probability contour and extract the errors. 

\begin{figure}[h]
\begin{center}
\includegraphics
[width=11cm, height=13cm]
{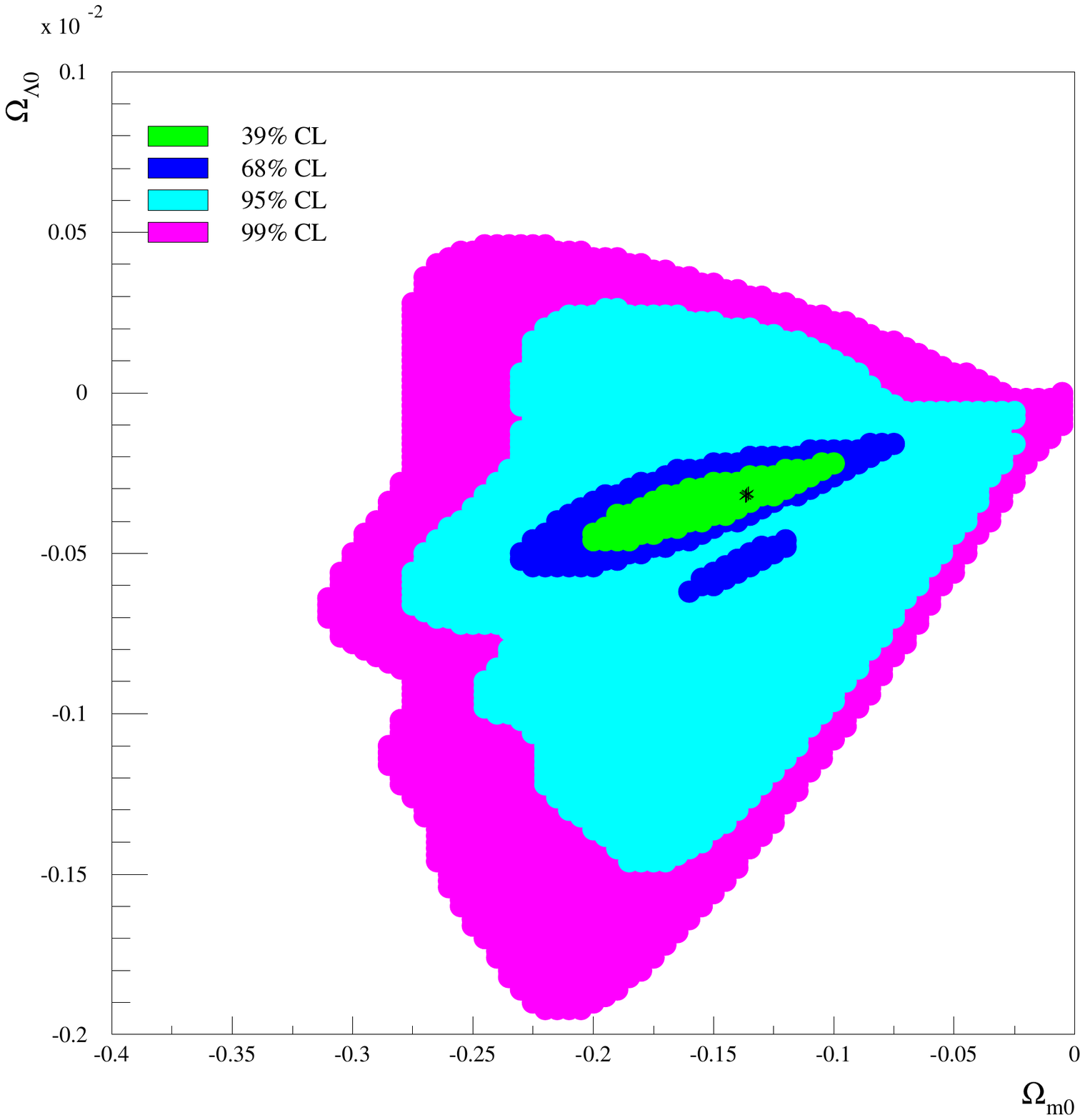}

\caption{$39\%$, $68\%$, $95\%$ and $99\%$ confidence level contours in the $\Omega_{m 0}$, $\Omega _{\Lambda 0}$ plane for exotic redshift.
The black star shows the minimum at $\Omega_{m0} = -0.15$ and $\Omega_{\Lambda 0} = -3.\, 10^{-4}$.}
\label{contour}
\end{center}
\end{figure}

Figure \ref{contour} shows the $39\%$, $68\%$, $95\%$ and $99\%$ confidence level contours in the $\Omega_{m 0}$, $\Omega _{\Lambda 0}$
plane. Smooth contours have been obtained by the use of a
Multi Layer Perceptrons (MLP) neural network \cite{frank,neural} with 2 hidden layers of 30 neurons each trained on
the results of the fit. Position of the minimum and errors are given in line 3 of table \ref{table1}. 

The cosmological constant is found to be compatible with zero at a level better than $10^{-3}$. As a first consequence, 
the exotic redshift is similar to the
standard one as it can be seen from equation (\ref{zsolve}). As a second consequence, the matter density becomes negative. 

This is a well known effect. Fitting a curved universe with vanishing cosmological constant to the Hubble diagram entails a negative matter density that emulates the recent acceleration of the expansion of our universe.

It is rather surprising that the pertubatively small effect of spin produces such drastic reductions of the cosmological constant (to zero) and of the matter density (to a negative value). Indeed, the fit of supernovae light curves attempts to minimize the effect
of the photon spin by suppressing the recent acceleration of the expansion of the universe, $\Lambda =0$. As a consequence, the preferred matter density becomes negative.

By comparing the $\chi^2$'s from line 1 and 3 in table \ref{table1} and using the log likelihood ratio statistical test hypothesis 
as in \cite{torsion} we
conclude that the exotic redshift is disfavored compared to the standard definition at a confidence level of at least $99.7\%$.

\section{Conclusions}

The confrontation of our exotic model with supernova data met with a -- to the best of our knowledge -- new challenge: in our model, the standardization of the absolute luminosities depends sensitively on the cosmological parameters. Therefore for every parameter choice we had to redo the global light curve fit for  every supernova with the SALT2 procedure and our analysis required three and a half CPU years. 

We had hoped that our exotic redshift would lower the mass density. It certainly did. However the obtained best fit for the exotic redshift is not good and its mass density negative. A negative mass density is not only physically problematic, it also is beyond the domain of validity of our perturbative solution, $Q+1>0$; a clean farewell to the exotic redshift, equation (\ref{exotic}).

Let us remark that our exotic redshift formula goes beyond the Etherington distance duality relation which is independent of the dynamics of the background geometry. Modifications of the Etherington relation due to weak gravitational birefringence have been considered by Schuller \& Werner \cite{sw}. These modifications can be confronted with data. We have not tried to test our modifications to the Etherington relation coming from the exotic redshift formula  because the latter is already ruled out by supernova data alone.

To end on a constructive note, we are presently trying to compute the birefringence induced on light while it passes through a gravitational wave, in the hope that this birefringence -- if it  exists -- might be detectable in interferometers with polarized laser beams.\\

\medskip\noindent
{\bf Acknowledgements:} The project leading to this publication has received funding from Excellence Initiative of Aix-Marseille University - A*MIDEX, a French ``Investissements d'Avenir'' programme.

\end{document}